\documentclass[longauth]{aa}
\usepackage{graphicx}
\usepackage{natbib}
\bibpunct{(}{)}{;}{a}{}{,} 

\include{references_rauer.bib}

\newcommand{\corot}{\emph{CoRoT}}

\newcommand{\corotp}{{CoRoT-5b}}
\newcommand{\corots}{{CoRoT-5}}

\newcommand{\RJ}{R$_{Jup}$}

\newcommand{\kms}{km\,s$^{-1}$}
\newcommand{\ms}{m\,s$^{-1}$}

\begin{document}

\title{Transiting exoplanets from the CoRoT space mission \\ VII. The ``hot-Jupiter"-type planet CoRoT-5b}

\author{Rauer, H.\inst{1, 2}
\and Queloz, D.\inst{3}
\and{Csizmadia, Sz.\inst{1}}
\and Deleuil, M.\inst{4}
\and Alonso, R.\inst{4}
\and Aigrain, S.\inst{5}
\and Almenara, J.M.\inst{6}
\and Auvergne, M.\inst{7}
\and Baglin, A.\inst{7}
\and Barge, P.\inst{4}
\and Bord$\acute{e}$, P.\inst{16}
\and Bouchy, F.\inst{8}
\and Bruntt, H.\inst{9}
\and Cabrera, J.\inst{1}
\and Carone, L.\inst{10}
\and Carpano, S.\inst{11}
\and De la Reza, R.\inst{19}
\and Deeg, H.J.\inst{6}
\and Dvorak, R.\inst{12}
\and Erikson, A.\inst{1}
\and Fridlund, M.\inst{11}
\and Gandolfi, D. \inst{14}
\and Gillon, M.\inst{3,18}
\and Guillot, T.\inst{13}
\and Guenther, E. \inst{14}
\and Hatzes, A.\inst{14}
\and H$\acute{e}$brard, G.\inst{8}
\and Kabath, P.\inst{1}
\and Jorda, L.\inst{4}
\and Lammer, H.\inst{15}
\and L$\acute{e}$ger, A.\inst{16}
\and Llebaria, A.\inst{4}
\and Magain, P.\inst{18}
\and Mazeh, T\inst{20}
\and Moutou, C.\inst{4}
\and Ollivier, M.\inst{16}
\and P$\ddot{a}$tzold, M.\inst{10}
\and Pont, F\inst{5}
\and Rabus, M. \inst{6}
\and Renner, S. \inst{1}
\and Rouan, D.\inst{7}
\and Shporer, A.\inst{20}
\and Samuel, B.\inst{16}
\and Schneider, J.\inst{17}
\and Triaud, A.H.M.J.\inst{3}
\and Wuchterl, G.\inst{14}}

\institute{Institute of Planetary Research, DLR, Rutherfordstr. 2, 12489
   Berlin, Germany
\and Center for Astronomy and Astrophysics, TU Berlin, Hardenbergstr. 36, 10623 Berlin, Germany
\and Observatoire de Gen\`eve, Universit´e de Gen\`eve, 51 Ch. des
   Maillettes, 1290 Sauverny, Switzerland
\and Laboratoire d'Astrophysique de Marseille, CNRS UMR 6110,
 Traverse du Siphon, 13376 Marseille, France
\and School of Physics, University of Exeter, Stocker Road, Exeter
   EX4 4QL, United Kingdom
\and Instituto de Astrofýsica de Canarias, E-38205 La Laguna,
   Tenerife, Spain
\and LESIA, CNRS UMR 8109, Observatoire de Paris, 5 place J.
    Janssen, 92195 Meudon, France
\and Institut d'Astrophysique de Paris, UMR7095 CNRS, Universit´e
   Pierre \& Marie Curie, 98bis Bd Arago, 75014 Paris, France
\and Sydney Institute for Astronomy, School of Physics, University of Sydney, NSW 2006, Australia
\and Rheinisches Institut f\"ur Umweltforschung an der Universit\"at zu
   K\"oln, Abt. Planetenforschung, Aachener Str. 209, 50931 K\"oln,
   Germany
\and Research and Scientific Support Department, European Space
   Agency, ESTEC, 2200 Noordwijk, The Netherlands
\and Institute for Astronomy, University of Vienna,
   T\"urkenschanzstrasse 17, 1180 Vienna, Austria
\and Observatoire de la C\^{o}te d'Azur, Laboratoire Cassiop\'{e}e, CNRS
   UMR 6202, BP 4229, 06304 Nice Cedex 4, France
\and Th\"uringer Landessternwarte Tautenburg, Sternwarte 5, 07778
   Tautenburg, Germany
\and Space Research Institute, Austrian Academy of Sciences,
   Schmiedlstrasse 6, 8042 Graz, Austria
\and Institut d'Astrophysique Spatiale, Universit\'{e} Paris XI, 91405
   Orsay, France
\and LUTH, Observatoire de Paris-Meudon, 5 place J. Janssen, 92195
    Meudon, France
\and Institut d'Astrophysique et de G\'{e}ophysique, Universit\'{e} de Li\`{e}ge,
  All\'{e}e du 6 aoˆut 17, Sart Tilman, Li\`{e}ge 1, Belgium
\and Observat´orio Nacional, Rio de Janeiro, RJ, Brazil
\and School of Physics and Astronomy, Raymond and Beverly Sackler
  Faculty of Exact Sciences, Tel Aviv University, Tel Aviv 69978,
  Israel
}

\date{00 / 00}

\abstract {} {The CoRoT  space mission continues to
photometrically monitor about 12000 stars in its field-of-view for a
series of target fields to search for transiting extrasolar planets ever
since 2007. Deep transit signals can be detected quickly in the
``alarm-mode" in parallel to the ongoing target field
monitoring. CoRoT's first planets have been detected in this mode.} {The CoRoT raw lightcurves are filtered for orbital residuals, outliers, and low-frequency stellar signals. The phase
folded lightcurve is used to fit the transit signal and derive the
main planetary parameters. Radial velocity follow-up observations
were initiated to secure the detection and to derive the planet mass.} {We report the detection of CoRoT-5b, detected
during observations of the LRa01 field, the first long-duration field in the galactic anti-center direction. CoRoT-5b is a ``hot Jupiter-type" planet with a radius of $1.388^{+0.046}_{-0.047}$  \RJ, a mass of $0.467^{+0.047}_{-0.024}$   M$_{Jup}$, and therefore,
a mean density of $0.217^{+0.031}_{-0.025}$   $\rm g\, cm^{-3}$. The planet orbits an F9V
star of 14.0 mag in 4.0378962$\pm$0.0000019 days at an orbital distance of $0.04947^{+0.00026}_{-0.00029}$ AU.} {}


\titlerunning{CoRoT-5b}
\maketitle

\footnote{Observations made with SOPHIE spectrograph at the Observatoire de Haute Provence (07B.PNP.MOUT), France, and HARPS spectrograph at ESO La Silla Observatory (072.C-0488(E), 082.C-0312(A)), and partly based on observations made at the
Anglo-Australian Telescope. The CoRoT space mission,
 launched on December 27, 2006, was developed and is operated
 by CNES, with the contribution of Austria, Belgium, Brasil, ESA,
 Germany, and Spain. }

\section{Introduction}

CoRoT started to search for the photometric signal of transiting
extrasolar planets in 2007, after its successful launch in December
2006, for details on the satellite see the pre-launch book \citep{2007baglin,Boisnard06} and \cite{auvergne2009}. The satellite
monitors about 12000 stars per exoplanet field-of-view in a series of short
($\sim$30 days) and long ($\sim$150 days) observing runs. Its magnitude range is 12 $\le$ m$_v$ $\le$ 16 mag. The
resulting stellar lightcurves are searched for periodic signals of
transiting extrasolar planets. Radial-velocity follow-up
measurements secure the  nature of the transiting body and allow us to derive
its mass.

The nominal lightcurve analysis for small transiting signals has
to await the completion of an observing run and detailed signal
analysis. The mission ``alarm-mode'' \citep{Quentin06,surace2008}, however, can be used to quickly trigger
follow-up measurements during ongoing observations of a target field. The
``alarm-mode" is used to increase the transmitted time-sampling for individual
stellar lightcurves in the CoRoT exoplanet channel. The sampling is increased from 512 sec to 32 sec if a transit-like signal is detected during the
observations. It therefore provides planetary
candidates early during an observing run, which are, however, biased towards
relatively large planetary candidates because of the limited data set available at this point.

CoRoT-5b is the fifth secured transiting planet detected by CoRoT. As CoRoT-1b to CoRoT-4b \citep{2008alonso,2008barge,deleuil2008,2008moutou,2008agrain}, it was first detected by the alarm-mode.  Here, we present the photometric detection of CoRoT-5b by the satellite based on pre-processed alarm-mode data,  the accompanying radial-velocity observations confirming its planetary nature, and the resulting planet parameters.

\section{Observations and data reduction}\label{observations}

CoRoT-5b was detected in the LRa01-field, the second long-run field of CoRoT. The field is located near the
anti-center direction of the galaxy at RA(2000): 06$^h$46$^m$53$^s$ and DEC(2000): -00$^\circ$12'00'' \citep{2006michel}. The observing sequence started on October 24, 2007 and finished after 112 days duration.
CoRoT observations usually have a very high duty cycle since data gaps are mainly caused by the regular crossings of the South Atlantic Anomaly (SAA), which typically last for about 10 min. During the observations of the LRa01 field, however, two longer interruptions occurred. An intermediate interruption of about 12 hours occurred eight days after the beginning of the observing run, and a longer data gap of about 3.5 days started on January 18, 2008, after a DPU reset. Finally, a duty cycle of 93 \% was achieved.

\begin{figure}
\includegraphics[width=\linewidth, height=6cm]{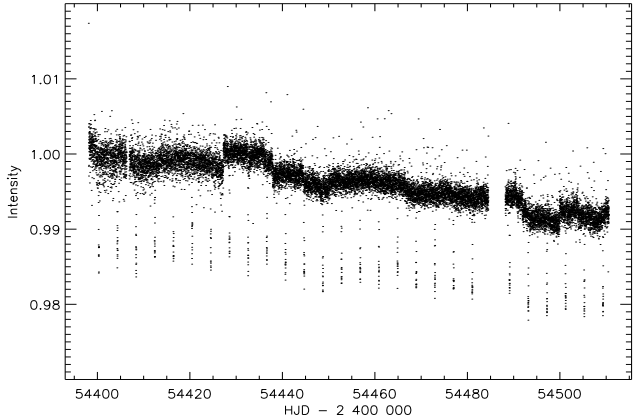}
\caption{Lightcurve of CoRoT-5 re-sampled to 512 sec time resolution. No corrections for data jumps due to ``hot pixels" have been applied in this figure to show the raw data quality. }\label{fig1_rauer}
\end{figure}

The alarm-mode was triggered after 29 days of observations.  When seven transit-like signals were detected, the  time sampling was switched to 32 sec. The alarm-mode data for CoRoT-5 are based on the analysis of ``white light" lightcurves, without using the color information of the CoRoT prism. In total 219,711 data points were obtained, 214,938 of it in oversampling mode. The data pipeline flags data points taken during the SAA crossing or affected by other events decreasing the data quality. When taking only unflagged data into account, the number of data points reduced to 204,092 in total and 199,917 as highly sampled.

The alarm-mode data were processed with a first version of the data reduction pipeline \citep{auvergne2009}. The pipeline corrects for the CCD zero offsets and gain, the sky background intensity and the telescope jitter. In addition, ``hot pixels" \citep{2008MNRAS.384.1337P} affect the lightcurves, causing sudden jumps in intensity of varying duration. The lightcurve of CoRoT-5 was, however, only moderately affected by such jumps, as can be seen in Fig. \ref{fig1_rauer}, which shows the full lightcurve. The oversampled part of the data set was re-binned to display the whole lightcurve with a 512 sec time sampling. The measured intensity decreases during the observing run, as observed for all stars in the fields.  Overall, CoRoT-5 only shows  a minor level of variability, without clear periodicity.

CoRoT measures stellar intensities by aperture photometry using optimized masks \citep{Llebaria:2003} that encompass the shape of the stellar point-spread-functions (PSFs). The bi-prism introduced in the light path of the exoplanet channel \citep{auvergne2009} causes relatively wide PSFs of unusual shapes that vary with e.g. stellar magnitude. Contaminating eclipsing binary stars within the PSF could mimic a planetary transit-like signal. Based on the pre-launch observations of the target field included in the {\sl Exo-Dat} data base \citep{deleuil2009}, the contamination of the mask of CoRoT-Exo5 is estimated to 8.4 \%. Refinement of this value will be performed in a more detailed future analysis using the dedicated windowing mask for this target star. We subtracted this flux level from the lightcurve before normalization to take  low level contamination into account.

The overall intensity trend and smaller scale variability of the lightcurve
were removed. To do this, we resampled the lightcurve to 512 seconds
sampling rate first and convolved this lightcurve with a fourth order Savitzky-Golay filter (similar to the treatment for CoRoT-2b \citep{2008alonso}). Then median averages were calculated for 24 hour segments of the lightcurve
(excluding the transit points and the data jumps), which was
fitted by a spline-curve. The original lightcurve was then divided by the
spline fit. The filtered lightcurve
was used for normalization and further analysis. The out-of-eclipse scatter of
CoRoT-5 was determined from the  standard deviation of data points in the
phase-folded lightcurve. It was found to be 0.0017 mag.

\begin{table}
\begin{center}
\caption{Radial velocity measurements of the star \corots\ obtained by SOPHIE and HARPS spectrographs from December 2007 to December 2008.}
\label{RVDATA}
\begin{tabular}{l l l} \hline  \hline
BJD & RV & Error\\
-2400000 & \kms & \kms \\
\hline
\multicolumn{3}{c}{SOPHIE}  \\
\hline
54463.4939000  & 48.947  & 0.017  \\
54465.5247100  & 48.816  & 0.028  \\
54506.3770000  & 48.767  & 0.016  \\
54525.3478500  & 48.860  & 0.020  \\
54528.2886100  & 48.933  & 0.031  \\
54544.3463300  & 48.925  & 0.026  \\
\hline
\multicolumn{3}{c}{HARPS}  \\
\hline

54548.583775  &    48.933  &  0.014  \\
54550.577783   &   48.792  &  0.021  \\
54551.584161   &   48.865  & 0.013  \\
54553.525234   &   48.883  & 0.010  \\
54554.546158   &   48.819  & 0.012  \\
54556.554191   &   48.929  & 0.010  \\
54768.852140   &   48.820  & 0.009  \\
54769.848137   &   48.900  & 0.008  \\
54771.850953   &   48.851  & 0.009  \\
54772.841289   &   48.827  & 0.010  \\
54773.847921   &   48.900  & 0.008  \\
54802.777527   &   48.929  & 0.009  \\
54805.748602   &   48.852  & 0.012  \\

\hline
\end{tabular}
\end{center}
\end{table}

\section{Photometric follow-up observation}

Photometric follow-up observations with higher spatial resolution than
CoRoT's (of $\approx$ 20\arcsec x 6\arcsec) are used to exclude the
presence of nearby contaminating eclipsing binaries (Deeg et al., this volume). Such observations of CoRoT-5 were performed at
the 80cm telescope at IAC, Tenerife, on the January 12, and 
March 11, 2008 at a spatial resolution of about 1.5\arcsec. These data
showed only one star bright enough to cause a potential false alarm,
about 8\arcsec\ southwest of the target. Observations obtained during
and out of a transit (``on/off photometry") showed, however, that this
contaminating star varies by less than 0.08 mag. This is far below the
variation of about 0.55 mag that is required to explain the observed signal in the CoRoT data.

\begin{figure}
\begin{center}
\includegraphics[height=8cm]{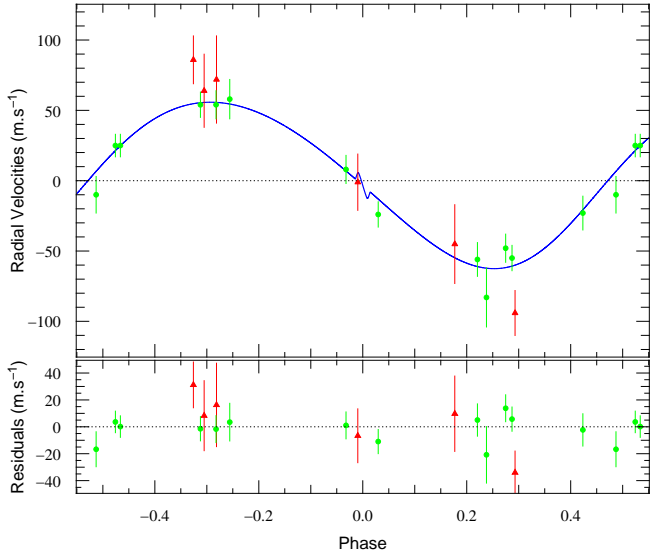}
\caption{Radial velocity measurements and Keplerian fit to the data including the Rossiter effect. Red: SOPHIE, green: HARPS.}\label{fig2_rauer}
\end{center}
\end{figure}

\begin{figure}
\begin{center}
\includegraphics[width=\linewidth,angle=0, width=9cm]{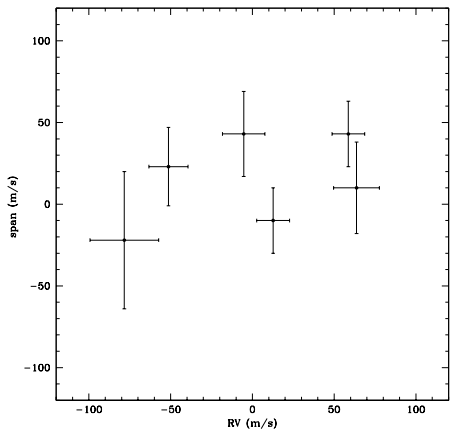}
\caption{Bisector analysis of CoRoT-5. }\label{fig3_rauer}
\end{center}
\end{figure}

\section{Radial velocity follow-up observations}

In January 2008,  after the identification of a  transit signal by the alarm-mode, \corots\ was observed  with the SOPHIE spectrograph installed on the 193 cm telescope at the  Haute Provence Observatory. Two  radial velocity measurements were taken at opposite  quadrature phases of the radial velocity variation expected from the transit ephemerides assuming  a circular orbit.  At this time the data were found to be compatible with  a  radial velocity amplitude suggesting  a Jupiter mass planet. Additional measurements  were obtained later in the season to confirm the reality of the signal but not enough to obtain a precise measurement of the orbit eccentricity. One year later, a new  series of measurements was obtained with the HARPS spectrograph installed on the 3.6m ESO  telescope at La Silla in Chile \citep{Mayor2003}. Both sets of data (SOPHIE and HARPS) have been processed as in \cite{bouchy2008}. Radial velocities (RV) were computed by weighted cross-correlation
\citep{baranne1996,pepe2005} with a numerical  G2-spectral template excluding spectral orders below 4200 \AA. Radial velocity values are listed in Table~\ref{RVDATA} and plotted in Fig. \ref{fig2_rauer}.

\begin{figure}
\begin{center}
\includegraphics[width=\linewidth,angle=90, height=3.5cm]{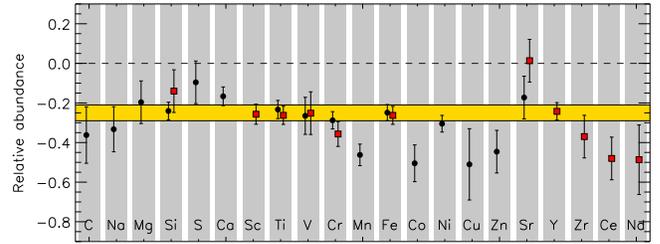}
\caption{Stellar abundances of CoRoT-5. Abundances found from neutral lines are marked by circles, for ionized lines box symbols are used. }\label{fig4_rauer}
\end{center}
\end{figure}

We analyzed the cross-correlation function computed from the HARPS spectra using the line-bisector technique according to the description in \cite{queloz2001} to detect possible spectral distortions caused by a faint background eclipsing binary mimicking a small RV amplitude signal. No correlation between the RV data and the bisector span was found at the level of the uncertainty on the data (Fig. \ref{fig3_rauer}).

The stability of the bisector, combined both with the amplitude of the
radial velocity and the accuracy of  transit lightcurve, is enough to
discard  an alternate  background eclipsing binary scenario.
In the case of a hypothetical background eclipsing binary, obtaining a sine-shaped radial-velocity signal would require
 a superimposed  spectrum  moving with the
same systemic velocity as the brightest component, and on an RV range
corresponding to the sum of the width of both CCF  line profiles.
This prerequisite constrains both on the mass of the potential
eclipsing  component and its companion. The example of HD41004
provides us with an interesting benchmark \citep{2002santos}.
This system was detected with a similar  radial velocity amplitude  but
with a  strong bisector correlation, and could be explained by a superimposed spectrum
with  3\% flux of the bright star. If one scales down this result
to  CoRoT-5, which has no bisector correlation, one finds that the
contrast ratio between the brightest star and the hypothetical
eclipsing binary is such that the eclipse must be very deep and the
radius of the eclipsing stars  much smaller than CoRoT-5. Considering
the quality of the CoRoT lightcurve such a binary scenario does not
match  the transit ingress and egress  timing and the detailed
shape of the curve.

\begin{table}
\begin{center}
\caption{Parameters of the parent star CoRoT-5.}\label{star}
\begin{tabular}{l l l} \hline \hline
parameter & value & source\\ \hline
 RA &   06$^h$ 45$^m$ 07$^s$ & {\sl Exo-Dat} \\
 DEC & 00$^\circ$ 48$\arcmin$ 55$\arcsec$ & {\sl Exo-Dat} \\
  epoch & 2000.0 & \\
  type & F9V & {\sl Exo-Dat,} \\
   & & {\sl AAOmega} \\
  $V$ & 14.0  & {\sl Exo-Dat} \\
  GSC2.3 ID &	N82O011953 & \\
 2MASS ID &	06450653+0048548 & \\
  $v \sin i$ [km s$^{-1}$] & 1$\pm1$\ & VWA  \\
  $\xi_t$ [km s$^{-1}$] & 0.91$\pm$0.09 & VWA\\
  $T_{\rm eff}$  [K] & 6100$\pm$65 & VWA\\
 $\log g$ & 4.19$\pm$0.03 & VWA\\
 $[M/H]$  & $-0.25 \pm 0.06$ & VWA\\
 M$_{star}$ [M$_\odot$] & 1.00$\pm$0.02 & Evolut. tracks\\
  R$_{star}$ [R$_{\odot}$] & 1.186$\pm$0.04 & Evolut. tracks\\
  $M^{(1/3)} / R$   [M$_\odot^{1/3}$ / R$_{\odot}$] & 0.843$\pm$0.024 & lightcurve \\
  age [Gyr]& 5.5 - 8.3 & photometry \\
                 &                & +Evolut. tracks \\
  \hline
\end{tabular}
\end{center}
\end{table}

\section{Properties of the central star}
 We determined the fundamental parameters of the host star carrying out a spectral analysis of the set of HARPS spectra acquired for radial velocity measurements. The individual spectra were reduced with the HARPS standard pipeline. The extracted spectra were corrected for  cosmics impacts, for the Earth and the stars velocity, and then corrected for the blaze function and normalized, order by order, to increase the signal-to-noise (S/N). The S/N level in the continuum is around 40 in the range 5\,000-6\,500\,\AA\ and it decreases to 15 towards the blue at 4\,000\,\AA.

Spectroscopic observations of the central star have also been performed in January 2008 with the AAOmega multi-object facility at the Anglo-Australian Observatory. By comparing the low-resolution (R=1300) AAOmega spectrum of the target with a grid of stellar templates, as described in \cite{2003frasca} and \cite{2008gandolfi}, we derived the spectral type and luminosity class of the star (F9 V).

As for the previous planet host stars, we used different methods to derive Corot-5 atmospheric parameters: line profile fitting with the SME \citep{1996valenti} and the VWA packages \citep{bruntt2002,bruntt08}. We find general agreement and here we quote the results from VWA. The star has a very low projected rotational velocity, $v \sin i = 1\pm1$\,km\,s$^{-1}$. More than 600 mostly non-blended lines were selected for analysis in the wavelength range 3\,990--6\,810 \AA.
VWA uses atmosphere models from the grid by \cite{heiter2002} and atomic parameters from the VALD database \citep{kupka1999}. The abundance determined for each line is computed relative to the result for the same line in the solar spectrum from \cite{hinkle2000}, following the approach of \cite{bruntt08}. The results for CoRoT-5 are shown in Table \ref{star}. Using these parameters for the atmospheric model, we determined the abundances of 21 individual elements.
 The uncertainty on the abundances includes a contribution of 0.04 dex due to the uncertainty on the fundamental parameters. The abundance pattern is shown in Fig.~\ref{fig4_rauer}. The overall metallicity is found as the mean abundance of the elements with at least 20 lines (Si, Ca, Ti, Cr, Fe, Ni) giving $[M/H]$\,$=-0.25\pm0.04$. We did not include Mn, as this has a significantly lower abundance. The metallicity and the 1-$\sigma$ error bar is indicated by the horizontal bar in Fig.~\ref{fig4_rauer}. There is no evidence of the host star being chemically peculiar, except Mn.

The fundamental parameters of the parent star, its mass and radius were
subsequently derived using stellar evolutionary tracks as presented in
\cite{deleuil2008} plotted in a M$^{(1/3)}$/R - $T_{\rm eff}$ HR diagram. The
stellar density parameter was derived from the lightcurve fitting (see sect.
7). We determined the mass and radius of the star to: M$_{\rm star}$ = 1.00$\pm$0.02 M$_\odot$
and  R$_{star}$ = 1.186$\pm$0.04 R$_\odot$. As a final
check, we calculated the corresponding surface gravity $\log g = 4.311\pm$0.033
 while the spectroscopic value is $4.19\pm0.03$. These two values of
$\log g$ are comparable with each other at the $3\sigma$ level. Based on our photometric analysis, we estimate
the age of the star to 5.5 - 8.3 Gyrs. The spectra show no sign of Ca II
emission or of a strong Li I absorption line, which is consistent with a relatively evolved star.

\begin{figure}
\includegraphics[width=\linewidth]{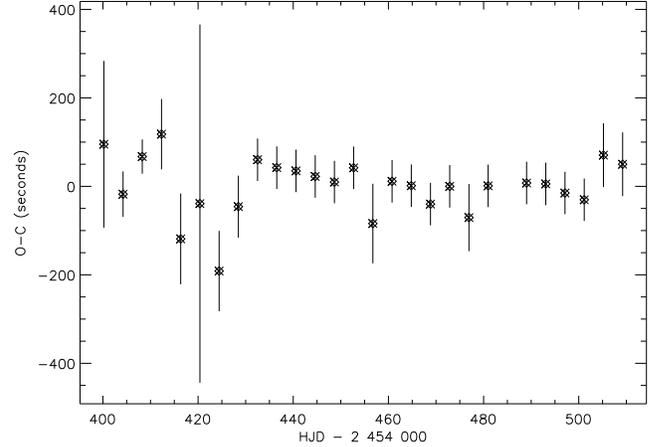}
\caption{The O-C diagram of the CoRoT-5b system. No clear period variation
can be seen. 
}\label{fig5_rauer}
\end{figure}

\section{Period determination and transit timing variations}

In total, 27 individual transit events are clearly seen, separated by an orbital period of about 4.03 days. One event was lost in a data gap.

First, we estimated the mid-times of each transit by applying the so-called Kwee-van Woerden method \citep{kvw56}. This method mirrors the lightcurve around a pre-selected time-point, T, computes the differences of original and mirrored lightcurves and then searches for an optimum T.  The $O-C$ diagram of the system was constructed, based on the resulting transit times and an initial guess of the period. A linear fit of this diagram yielded an improved estimate of the period. This period value was then refined with the following procedure. The lightcurve was phase-folded using this previously determined period and then averaged. The size of the bin used was 0.001 in phase (or to 5.81 minutes, using the final period). Then, this lightcurve was fitted (see the next section) by  a theoretical transit lightcurve. The transit mid-times were then determined again by cross-correlating the observed and the theoretical lightcurve. This resulted in more precise mid-times of the transit and a new $O-C$ curve. Another linear fit to this $O-C$ diagram yielded a better period value, and the whole procedure was repeated. The final O-C diagram can be seen in Fig. \ref{fig5_rauer}. The resulting ephemeris is given in Table \ref{planettable}.

There is no obvious period variation present in the $O-C$ diagram. The first part of the lightcurve was obtained with the 512 sec sampling rate, so the first seven minima  typically consist of only 20 data points. Thus, they have larger scatter and uncertainties. The next twenty minima were obtained with the high sampling rate (32 sec) and typically consist of a few hundred data points, leading to  much higher accuracy. If one takes  only these high-resolution minima into account, the constancy of the period is clearer. However, we cannot exclude that small period variations are present in the system. The upper limit of such a period variation was estimated by a quadratic fit to the data, which showed that it should be less than 0.42 seconds/cycle.

\section{Analysis of  parameters of \corotp}

\begin{figure}
\includegraphics[width=\linewidth]{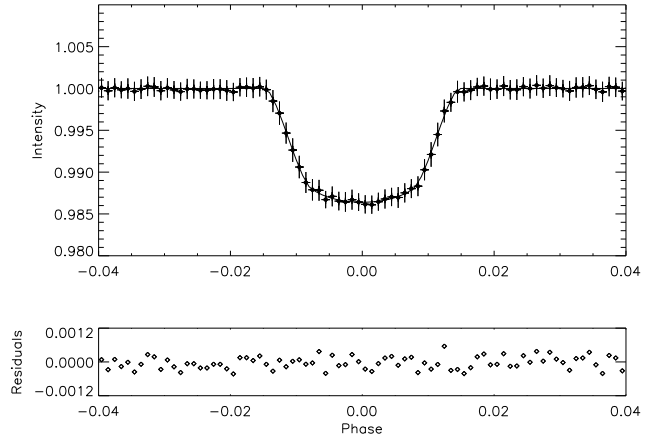}
\caption{Top: Phase-folded lightcurve of CoRoT-5b. Bottom: Residuals of fitted transit curve.}\label{fig6_rauer}
\end{figure}

The final phase-folded lightcurve of the transit event is seen in Fig.
\ref{fig6_rauer}. The transit signal shows a depth of about 1.4 \% and lasts
for about 2.7 hours. We derived the planetary parameters by fitting
simultaneously the lightcurve of \corots\  with the SOPHIE and HARPS radial
velocities. A planetary model on a Keplerian orbit in the formalism of
\cite{Gimenez2006a} and \cite{gimenez2006ApJ} was fitted to the data using a
Markov Chain Monte-Carlo (MCMC) code described in Triaud et al. (in prep.) but
using $e.\cos\omega$ and $e.\sin\omega$  instead of $e$ and $w$ as free
parameters for better error estimation. In the fit  a quadratic limb-darkening
law was assumed at $u_+ =0.616$ and $u_- =0$. In the initial \textit{burn-in}
phase of the MCMC adjustment, 15,000 steps were chosen  to allow the fit to
converge. A further 50,000 steps were  used to derive the best parameters and
their errors. In the fit, there are eight fitted parameters plus two $\gamma$
velocities and a normalization factor, totalling 11 free parameters. In
addition, the fit assumed the presence of a Rossiter-McLaughlin effect with the two
fixed parameters $v sin\,i = 1.0$  km s$^{-1}$ and $\lambda = 0$ ($\lambda$:
angle between stellar rotation axis and normal vector of the orbital plane). A
Bayesian penalty is added to the $\chi^2$ creating a prior for $M_\star = 0.99
\pm 0.02$.  The fit to the rv measurements is shown in Fig. \ref{fig2_rauer},
and the derived fitting parameters are shown in Table \ref{RVfittable}.

\begin{table}
\begin{center}
\caption{Parameters of the \corots\ system derived from the combined MCMC analysis. }\label{RVfittable}
\begin{tabular}{l l l} \hline  \hline
Fitted Parameters & Value & Units \\
\hline\\
 ${(R_p/R_{star})^2}$& $0.01461^{ +0.00030}_{-0.00032}$& \\[1mm]
 $t_T$  &$0.0290 ^{+0.00038} _{ -0.00053}$& \\[1mm]
 $b$ & $0.755^{+  0.017}_ {-0.022}$& \\[1mm]
 $K$  &$59.1^{+6.2}_{-3.1}$ & \ms \\[1mm]
$e \cos\omega$ &$-0.057^{+0.048}_{-0.020}$& \\[1mm]
$e \sin\omega$ &$-0.071^{+0.147} _{-0.130}$& \\[1mm]
\hline
\end{tabular}
\begin{list}{}{}
 $t_T$ denotes the transit duration given in fraction of phase, $b$ the impact parameter and $K$ the RV semi-amplitude.
\end{list}
\end{center}
\end{table}

\begin{table*}
\begin{center}
\caption{The derived planet parameters.}\label{planettable}
\begin{tabular}{l l l} \hline  \hline
Derived physical  parameters & Value & Units \\
\hline\\
Transit epoch T$_0$ &$2454400.19885\pm0.0002$ & HJD\\[1mm]
Orbital period $P$  & $4.0378962\pm0.0000019$ & days \\[1mm]
Orbital semi-major axis $a$ & $0.04947^{+0.00026}_{-0.00029}$ & AU\\[1mm]
Orbital inclination $  i$  &$ 85.83^{+0.99}_{-1.38}$ & degrees\\[1mm]
Orbital eccentricity $e$  & $ 0.09^{+0.09}_{-0.04} $& \\[1mm]
Argument of periastron $\omega$  &$-2.24^{+5.05}_{-0.84}$ & rad\\[3mm]
Planet radius $R_\textrm{\scriptsize p}$ & $1.388^{+0.046}_{-0.047}$  & $R_\textrm{\scriptsize J}$\\[1mm]
Planet mass $M_\textrm{\scriptsize p}$  &$0.467^{+0.047}_{-0.024}$  & $M_\textrm{\scriptsize J}$\\[1mm]
Mean planet density $\rho_\textrm{\scriptsize p}$  & $0.217^{+0.031}_{-0.025}$   & $g cm^{-3}$ \\[1mm]
Planetary surface gravity $\log g_\textrm{\scriptsize p}$¤  & $7.77^{+0.14}_{-0.08}$ & cgs \\[1mm]
Zero albedo equilibrium temperature $T_{eq}$ & \textbf{$1438\pm39$} & K \\[1mm]
\hline
\end{tabular}
\end{center}
\end{table*}

In addition, a model transit curve \citep{2002mandel} was fitted to the
photometric phase folded transit curve separately. The parameters fitted are
the center of transit, the planet radius expressed in stellar radii, the
semi-major axis in stellar radii and the orbital inclination. In this fit the
limb-darkening coefficients (u$_1$ and u$_2$) were free parameters, assuming a
quadratic limb-darkening law. The fitting method follows a Metropolis-Hastings
algorithm, which is a kind of Markov Chain Monte-Carlo procedure. The fitting
procedure was performed ten times with different starting values to find the
global minimum in $\chi^2$.  The errors of the fit were estimated from the
standard deviations of the points in the chain. In addition to the transit
curve, a third light component is included as a free parameter in the fit. In
this way, we could check whether another contaminant is present, which remained
unresolved in the photometric follow-up. However, no such additional source of
light was found. The transformation between contamination factor $c$ and
the third light $l_3$ is $c = l_3/(1-l_3)$. We had $c=0.005 \pm 0.024$. Since
we already removed the known contaminant factor from the lightcurve  (see
Section 2), we could therefore conclude that no further observable contaminant is present
in the lightcurve of CoRoT-5. The planet parameters derived from this fit
agree with the simultaneous fitting within the error bars, so we do not
 report them again here.

The resulting planetary parameters based on the MCMC approach with fixed
limb-darkening coefficients and without any third light are summarized in
Table \ref{planettable}. The major uncertainties on the planet are, as usual,
introduced mainly from the uncertainty of the stellar parameters.

\section{Summary}

We report the discovery of a ``hot-Jupiter-type" planet, CoRoT-5b, orbiting
a type F9V star of 14.0 mag.
The planet mass and radius  were derived to $0.467^{+0.047}_{-0.024}$ M$_{Jup}$
and $1.388^{+0.046}_{-0.047}$   R$_{Jup}$, respectively.  It orbits its central
star at $0.04947^{+0.00026}_{-0.00029}$ AU orbital distance. The determined
eccentricity is low (see Table \ref{planettable}), but further radial
velocity measurements would be needed for a more accurate determination.

CoRoT-5b has a density of $ 0.217^{+0.031}_{-0.025}$ g cm$^{-3}$, similar to the
planets WASP-12b and WASP-15b \citep{2009hebb,west2009}, implying that it belongs to
the planets with the lowest mean density found so far.
As such, it is found to be larger by 20\% than standard evolution
models \citep{guillot2006} would predict. Standard recipes that
account for missing physics (kinetic energy transport or increased
opacities) can explain this large size, and predict that the planet is
mostly made of hydrogen-helium, with at most $28\,\rm M_\oplus$ of
heavy elements (maximum value obtained in the kinetic energy model,
assuming 0.5\% of the incoming energy is dissipated at the planet
center). Thus, CoRoT-5b supports the proposed link between the metallicity of planets and of their host star.


\bigskip

\textbf{Acknowledgements}

HJD and JMA acknowledge support from grant ESP2007-65480-C02-02 of the
Spanish Education and Science Ministry. Some of the data published in
this article were acquired with the IAC80 telescope operated by the
Instituto de Astrof\'\i sica de Tenerife at the Observatorio del
Teide. The German CoRoT Team (TLS and Univ. Cologne) acknowledges
DLR grants 50OW0204, 50OW0603, 50QP07011. RA acknowledges support by grant  CNES-COROT-070879.
The building of the input \corot/Exoplanet catalog was made possible by observations collected for years at the Isaac Newton Telescope (INT), operated on the island of La Palma by
the Isaac Newton group in the Spanish Observatorio del Roque de Los
Muchachos of the Instituto de Astrofisica de Canarias.

\bigskip

\bibliographystyle{aa} 

\bibliography{11902references}

\end{document}